\documentclass[11pt]{article}
\usepackage[dvips]{graphicx}
\def\be{\begin{equation}}
\def\ee{\end{equation}}
\def\bea{\begin{eqnarray}}
\def\eea{\end{eqnarray}}
\def\l#1{\overline{#1}}
\def\b#1{\beta_{#1}}
\def\g#1{\gamma_{#1}}
\def\du{\mu\frac{\partial}{\partial\mu}}
\def\f#1#2{\frac{#1}{#2}}
\def\gb{\f{\g{0}}{2\b{0}}}
\newcommand{\partialslash}{\partial \!\!\! /}

\newcommand{\sect}[1]{ \section{#1} \setcounter{equation}{0} }
\begin{document}

\hfill LTH--534

{\begin{center} {\LARGE {Dynamical mass generation by source
inversion:
calculating the mass gap of the chiral Gross-Neveu model} } \\
[8mm] {\large K. Van Acoleyen$^a$, J.A. Gracey$^b$ and H.
Verschelde$^a$}
\end{center}
}

\begin{itemize}
\item[$^a$] Department of Mathematical Physics and Astronomy, University of
Ghent, Krijgslaan 281 (S9), 9000 Ghent, Belgium.
\item[$^b$] Theoretical Physics Division, Department of Mathematical Sciences,
\\
University of Liverpool, Liverpool, L69 7ZF, United Kingdom.
\end{itemize}

\vspace{5cm} \noindent {\bf Abstract.} We probe the $U(N)$ chiral
Gross-Neveu model with a source-term $J\l{\Psi}\Psi$. We find an
expression for the renormalization scheme and scale invariant
source $\widehat{J}$, as a function of the generated mass gap. The
expansion of this function is organized in such a way that all
scheme and scale dependence is reduced to one single parameter
$d$. We obtain a non-perturbative mass gap as the solution of
$\widehat{J}=0$. A physical choice for $d$ gives good results for
$N>2$. The self-consistent minimal sensitivity condition gives a
slight improvement.

\newpage
\section{Introduction}

In a previous paper, we developed a method for dynamical mass
generation in asymptotically free quantum field theories. It was
applied to the ordinary Gross-Neveu model, \cite{gn74}, and a mass
gap was found, \cite{vv01}, which agreed very well with the exact
result, \cite{fnw91}. In this paper we will apply the same method
to the non-abelian Thirring model (NATM) or chiral Gross-Neveu
model (CGNM), \cite{gn74}. It is another one of those rare quantum
field theories where exact results, like the mass gap, can be
obtained. In \cite{fnn92} the mass gap is calculated exactly in
terms of $\Lambda$ which is the non-perturbative mass parameter
which sets the scale for the running coupling in a certain scheme.
Comparing our results with \cite{fnn92} will provide another check
on the accuracy of our method.

The idea behind the method is very simple. A source term,
$J\l{\Psi}\Psi$, is added to the NATM-Lagrangian and then we
calculate the mass gap using ordinary perturbation theory to
obtain the perturbative expansion for $m(J)$. As a consequence of
asymptotic freedom, this expansion is only valid for large values
of $J$. If we let $J$ approach zero, the coupling constant grows
too large, and the perturbative expansion for $m(J)$ becomes
invalid. Therefore, we cannot take the limit $J\rightarrow 0$. If
instead we consider the perturbative expansion for the inverted
relation $J(m)$, perturbation theory remains valid in the limit
$J\rightarrow 0$, provided that a solution $m$ exists for
$J(m)=0$, which is not too small. As in ordinary perturbation
theory, the result for the mass gap $m$ is renormalization scheme
(RS) and scale dependent. To eliminate the mass renormalization
dependence we use the scheme and scale independent quantity
$\widehat{J}$, instead of $J$. Exchanging $g^2(\mu)$ for
$1/(\b{0}\ln\f{\mu^2}{\Lambda^2})$ as the expansion parameter,
reduces the remaining dependence to one single number $d$, which
can be fixed by some external physical condition, or in a more
self-consistent approach, by the principle of minimal sensitivity.

The paper is organized as follows. In section \ref{method} we will
present the results necessary for application of the source
inversion for the NATM. In order to avoid unnecessary repetition,
we shall refer to the paper \cite{vv01} for the derivation of the
general formula. The outcome of our calculations will be discussed
in section \ref{numerical}. As a bonus we will show that
reparametrization of the $d$-dependence will enable us to solve
the mass gap equation exactly. Details of the exact evaluation of
the finite parts of the two loop Feynman integrals which occur in
the sunset topology are given in an appendix.

\section{The non-abelian Thirring model}\label{method}
The $U(N)$ invariant NATM describes the interaction of $N$ single
flavor Dirac fermions $\Psi_{a}$, $a=1,\ldots,N$ in two dimensions
with the (massless) Lagrangian \be \mathcal{L} ~=~
i\overline{\Psi} \partialslash\Psi ~-~
\frac{1}{2}g^{2}{(\overline{\Psi}\gamma^{\mu}T^{i}\Psi)}^{2}
\label{natm} \ee where $T^{i}$, $i=1,\ldots, N^2-1$, are the
generators of $SU(N)$ with the normalization
$\mbox{Tr}(T^{i}T^{j})=\f{1}{2}\delta^{ij}$. (Note that our
coupling constant $g^2$ is two times the coupling constant $g^2$
of \cite{fnn92}.) This model is also known as the CGNM because a
Fierz-transformation of the interaction term leads to the
equivalent Lagrangian \be \mathcal{L} ~=~ i\overline{\Psi}
\partialslash\Psi ~+~ \frac{g^2}{4} \left(
{(\overline{\Psi}\Psi)}^{2} ~-~
{(\overline{\Psi}\gamma^{5}\Psi)}^{2} \right) ~+~
\frac{g^2}{4N}{(\overline{\Psi}\gamma^{\mu}\Psi)}^{2} ~. \ee The
NATM is asymptotically free, \cite{df73} and possesses, apart from
the $U(N)$ invariance, a chiral $U(1)$ symmetry. In ordinary
perturbation theory this symmetry remains unbroken and no mass gap
is generated.

We begin by perturbing (\ref{natm}) with a $\l{\Psi}\Psi$
composite operator to produce the new Lagrangian
\be\mathcal{L}_{J} ~=~ i\overline{\Psi} \partialslash\Psi ~-~
J\l{\Psi}\Psi ~-~ \frac{g^2}{2} {(\overline{\Psi}\gamma^{\mu}
T^{i}\Psi)}^{2} ~. \label{natmj} \ee The detailed three loop
renormalization of this model using dimensional regularization has
been given in \cite{vvdk96,bg99} which built on the one and two
loop calculations of \cite{df73,de88,bcpr89,vvdk96}. The results
for the $\beta$ and $\gamma$-functions in the $\l{MS}$-scheme are
$(g^2=g^2(\mu))$: \bea \left. \du  J
\right|_{J_{0},g_{0},\epsilon}&\equiv& -~ J\gamma(g^{2}) ~\equiv~
-~ J(\g{0}g^{2} + \g{1}g^{4} + \g{2}g^{6} + \ldots)
\nonumber \\
&=& -~ J \left( \f{(N^2-1)}{2\pi
N}g^2+\f{(N^2-1)(N-4)}{16{\pi}^2N}g^4 \right.
\nonumber\\
&&\left.+~\f{(N^2-1)(16N^2-12N^3+3N^4+5N^2-26)}{128
{\pi}^3N^3}g^{6}+\ldots \right) \nonumber \\
\label{gamma} \eea and \bea \left. \du g^{2}
\right|_{g_{0},\epsilon} &\equiv& \beta(g^{2}) ~=~ -~
2(\b{0}g^{4}+\b{1}g^{6}+\b{2}g^{8}+\ldots)\nonumber\\
&=& -\, 2\left( \! \f{N}{4\pi}g^4-\f{N}{8{\pi}^2}g^6 +
\f{5N+\f{3}{2}N^3-\f{11}{2}N+\f{39}{2}\f{1}{N}}{64{\pi}^3}g^{8}+\ldots\right)~.
\nonumber \\
\label{beta} \eea To apply the source inversion at two loop order
we also require the two loop perturbative result for the mass gap.
To determine this we have computed the fermion two point-function
at two loops and extracted the finite part exactly after
performing the renormalization. The values for the integrals we
obtained have been checked against the numerical results of
\cite{agikn97} for the mass gap of the ordinary Gross Neveu model.
We find \bea m(J) &=& J \left[ 1 - \f{(N^2-1)}{N}
\Big(\ln\f{J^2}{\mu^2}+1\Big)
\f{g^2}{4\pi} \right. \nonumber \\
&& \left. ~~~+~ \f{(N^2-1)}{2N} \left( 2\zeta(2)
\left(3-\f{1}{N}\right)
+ \f{(3N^2-2N-4)}{N} \right. \right. \nonumber \\
&& \left. \left. ~~~+~ \f{(7N^2+4N-6)}{N} \ln\f{J^2}{\mu^2} +
\f{(2N^2-1)}{N} \ln^2\f{J^2}{\mu^2} \right) \f{g^4}{16\pi^2}
+\ldots\right]  \nonumber \\
\eea where $\zeta(n)$ is the Riemann zeta function. From this one
easily arrives at the expansion for the inverted relation \bea
J(m) &=& m \left[1+\f{(N^2-1)}{N}\Big(\ln\f{m^2}{\mu^2}+1\Big)
\f{g^2}{4\pi}+\f{(N^2-1)}{2N} \left( 2\zeta(2)
\left(\f{1}{N}-3\right) \right. \right.
\nonumber\\
&& \left. \left. ~~~~+~ \f{(3N^2+2N-2)}{N} + \f{(N^2-4N-2)}{N}
\ln\f{m^2}{\mu^2} \right. \right. \nonumber \\
&& \left. \left. ~~~~-~ \f{1}{N} \ln^2\f{m^2}{\mu^2} \right)
\f{g^4}{16\pi^2} +\ldots \right] ~. \label{jm} \eea We define
$X_{0}$ and $Y_{0}$ as the coefficients which do not multiply a
logarithm \be J(m) ~\equiv~ m(1+g^2(m)X_{0}+g^4(m)Y_{0}+\ldots) ~.
\nonumber \ee The expansion (\ref{jm}) is highly scheme and scale
dependent. This dependence is reduced drastically if we replace
$J(\mu)$ with $\widehat{J}$ which is the scheme and scale
independent quantity associated with $J$, and then expand in
powers of $1/\left(\b{0}\ln\f{\mu^{2}}{\Lambda^{2}}\right)$ rather
than in $g^{2}(\mu)$. Starting with the expansion for $J(m)$ in a
general scheme, we found \cite{vv01} \bea
\widehat{J}&=&m\left(\b{0}\ln
\f{m^{2}}{{\Lambda^2_{\l{MS}}}}+d\right)^{\f{\g{0}}{2\b{0}}}\times\nonumber\\
&&\left[1 +\f{1}{\left(\b{0}\ln
\f{m^{2}}{{\Lambda^2_{\l{MS}}}}+d\right)}
\bigg[A_{0}+\f{\g{0}\b{1}}{2{\b{0}}^{2}} \ln\left(\ln
\f{m^{2}}{{\Lambda^2_{\l{MS}}}}+\f{d}{\b{0}}\right)
-\f{d\g{0}}{2\b{0}}\bigg]\right.\nonumber\\
&&\left.+~\f{1}{\left(\b{0}\ln
\f{m^{2}}{{\Lambda^2_{\l{MS}}}}+d\right)^{2}}
\bigg[B_{0}+A_{0}\left(\gb-1\right)\f{\b{1}}{\b{0}} \ln\left(\ln
\f{m^{2}}{{\Lambda^2_{\l{MS}}}}+\f{d}{\b{0}}\right) \right.
\nonumber \\
&&\left.+~\f{\b{1}^2}{\b{0}^2}\left(\left[\ln\left(\ln
\f{m^{2}}{{\Lambda^2_{\l{MS}}}}+\f{d}{\b{0}}\right)\right]^{2}
\gb\left(\f{\g{0}}{4\b{0}}-\f{1}{2}\right) \right. \right. \nonumber \\
&& \left. \left. +\gb\ln\left(\ln
\f{m^{2}}{{\Lambda^2_{\l{MS}}}}+\f{d}{\b{0}}\right)\right)\right.\nonumber\\
&&\left.-~\gb\left(\f{\b{2}}{\b{0}}-\f{{\b{1}}^{2}}{{\b{0}}^{2}}\right)+d^{2}
\left(\f{\g{0}}{4\b{0}}\left(\gb-1\right)\right)\right.\nonumber\\
&&\left.+~d\left(A_{0}\left(1-\gb\right)-\f{\g{0}\b{1}}{2{\b{0}}^{2}}
\right.
\right.\nonumber \\
&&\left.\left. +~\f{\g{0}\b{1}}{2{\b{0}}^{2}}\ln\left(\ln
\f{m^{2}}{{\Lambda^2_{\l{MS}}}}+\f{d}{\b{0}}\right)\left(1-\gb\right)
\right)\bigg] + \mathcal{O}\left(\f{1}{\b{0}\ln
\f{m^{2}}{{\Lambda^2_{\l{MS}}}}+d}\right)^{\!3} \right] \nonumber \\
\eea with \bea A_{0} &\equiv& X_{0} ~-~
\f{1}{2}\left(\f{\g{1}}{\b{0}}
-\f{\g{0}\b{1}}{\b{0}^{2}}\right)\nonumber\\
B_{0}&\equiv&\f{X_{0}}{2}\left(\f{\g{0}\b{1}}{\b{0}^{2}}-\f{\g{1}}{\b{0}}
\right)- \f{\g{2}}{4\b{0}}+\f{\g{1}\b{1}}{4\b{0}^{2}}-
\f{\g{0}}{4\b{0}}\left(\f{\b{1}}{\b{0}}\right)^{2}+\f{\g{0}\b{2}}{4\b{0}^2}
+\f{\g{1}^{2}}{8\b{0}^{2}}\nonumber\\
&&-\f{\g{1}\g{0}\b{1}}{4\b{0}^{3}}+\f{\g{0}^{2}\b{1}^{2}}{8\b{0}^{4}}+
Y_{0} ~. \eea All the scheme and scale dependence now resides in
$d$ $\equiv
\b{0}\ln\Big(\f{\Lambda^2_{\l{MS}}}{{\Lambda}^2}\f{\mu^2}{m^2}\Big)$
and we can recover the original NATM, by putting the naked source
$J_{0}$ equal to zero \be J_{0}(m) ~\sim~ \widehat{J}(m) ~=~ 0 ~.
\ee We find a non-perturbative mass gap which is a solution of
\bea 1+\f{1}{(\b{0}\ln
\f{m^{2}}{{\Lambda^2_{\l{MS}}}}+d)}\Big[\ldots\Big]
+\f{1}{(\b{0}\ln
\f{m^{2}}{{\Lambda^2_{\l{MS}}}}+d)^{2}}\Big[\ldots\Big]+\ldots=
0~.\label{j=0} \eea The total series is of course $d$-independent
but one can only calculate it up to a certain order in
perturbation theory which will give us a mass gap that depends on
$d$. One can check that the $d$-dependence of the order $n$
truncated series is $\mathcal{O}\Big(\f{1}{\b{0}\ln
\f{m^{2}}{{\Lambda^2_{\l{MS}}}}+d}\Big)^{n+1}$. We will consider
two possible ways of fixing $d$. The first one reduces to a choice
for $\Lambda$, that corresponds to a physical scheme. The second
one fixes $d$ by the principle of minimal sensitivity. In
\cite{vv01} we used the value of the expansion parameter
$1/\left({\b{0}\ln \f{m^{2}}{{\Lambda^2_{\l{MS}}}}+d}\right)$ as a
source of error estimation. This works if the coefficients are of
order one. Assuming that the series is asymptotic, a rather large
value of the expansion parameter can still give reasonable
results, as long as the complete terms in the series are small. In
the next section we will show that this is indeed the case. For
the $2$-loop results it is better to estimate the error from the
second order term, than from the expansion parameter.

\section{Numerical results}\label{numerical} The exact result for
the mass gap was obtained in \cite{fnn92}, \be m ~=~
\f{e^{\f{1}{2N}}}{\Gamma\left(1-\f{1}{N}\right)}\Lambda_{PV} ~,
\ee where $\Lambda_{PV}$ is defined as the scale parameter for the
running coupling, with a condition on the normalized four point
function, calculated with a Pauli-Villars regularization. To
obtain $m/\Lambda_{\l{MS}}$ we need to determine the relationship
between the renormalized coupling $g$ of the dimensional
regularization $\l{MS}$-scheme and the coupling $g_{PV}$ used in
the Pauli-Villars scheme. This can be achieved by comparing the
normalized fermion four-point function to one loop order in both
schemes. We find \be g^2 ~=~
{g_{PV}}^2\left[1+\f{{g_{PV}}^2}{4\pi} \left(\f{N}{2}+1 \right)
+\ldots \right] \ee and hence \be \Lambda_{PV} ~=~
\Lambda_{\l{MS}}e^{-(\f{1+\f{N}{2}}{2N})} ~. \ee (See, for
example, \cite{cg79}.) So we finally arrive at \be m ~=~
\f{e^{-\f{1}{4}}}{\Gamma(1-\f{1}{N})}\Lambda_{\l{MS}} ~=~
e^{-\f{1}{4}}\Lambda_{\l{MS}} \left[ 1-\f{\gamma}{N}+\mathcal{O}
\left(\f{1}{N}\right)^2 \right] \ee where $\gamma$ is the
Euler-Mascheroni constant. \subsection{Physical scheme}As in
\cite{vv01}, we will define a \textit{physical} RS on the
normalized 4-point function of $\mathcal{L}_{J}$, to obtain a
physical value $d_{f}$ for $d$. Demanding that $g_{f}^2$ coincides
with the 4-point function at zero external momentum (see equation
(5.5) of \cite{bg99}), one arrives at \be g^2 ~=~ g_{f}^2
\left(1+\f{3N}{8\pi}g_{f}^2+\ldots \right) \ee Taking $\mu^2=m^2$
leads to $d_{f}=\f{3N}{8\pi}$.\footnote{As in \cite{vv01}, one can
show that every physical value for $d$ $(\sim N)$ gives the
correct $N\rightarrow \infty$ limit.}
\begin{table}\label{physical}
\begin{center}
\begin{tabular}{|r|r|r|r|r|r|r|}
\hline $N$&$m_{1}$&$m_{2}$&$y/\pi$&II&$N=\infty$&1/N \\
\hline 2&/&46.8\%&3.2&-0.37&77.2\%&26.1\%\\
3&/&10.3\%&2.3&-0.02&35.4\%&9.4\%\\
4&15.1\%&5.3\%&1.4&0.10&22.5\%&4.9\%\\
5&12.6\%&5.4\%&1.0&0.09&16.4\%&3.0\%\\
6&10.6\%&5.5\%&0.8&0.07&12.9\%&2.0\%\\
7&9.0\%&5.3\%&0.6&0.05&10.6\%&1.5\%\\
8&7.8\%&5.0\%&0.5&0.04&9.0\%&1.1\%\\
9&6.9\%&4.7\%&0.5&0.04&7.8\%&0.9\%\\
10&6.2\%&4.4\%&0.4&0.03&6.9\%&0.7\%\\
$\infty$&0\%&0\%&0&0&0\%&0\%\\ \hline
\end{tabular}
\caption{Physical scheme results.}
\end{center}
\end{table}
We now find the one and two loop mass gaps $m_{f1}$ and $m_{f2}$
as the solutions of the one and two loop truncation of (\ref{j=0})
with $d=d_{f}$. The deviations from the exact result for $m_{f1}$
and $m_{f2}$ for the $N\rightarrow \infty$ limit and the first
order result in $1/N$ have been displayed in table \ref{physical}
in terms of a percentage. We also provided the value of the two
loop expansion parameter $1/(2\b{0}\ln
\f{m_{f2}}{\Lambda_{\l{MS}}}+d_{f})\equiv y$ and the second order
term II of the series in (\ref{j=0}). For $N=2,3$ we find no
one-loop mass gap. All the other results lie somewhere between the
$N\rightarrow \infty$ and $1/N$ approximations. We also observe
convergence. In other words the comparison with the exact result
improves for the two loop truncation. From the $y/\pi$ and the II
columns we learn that II clearly gives a better indication on the
size of the error.

\subsection{Minimal sensitivity}
\begin{figure}
\begin{center}
\includegraphics{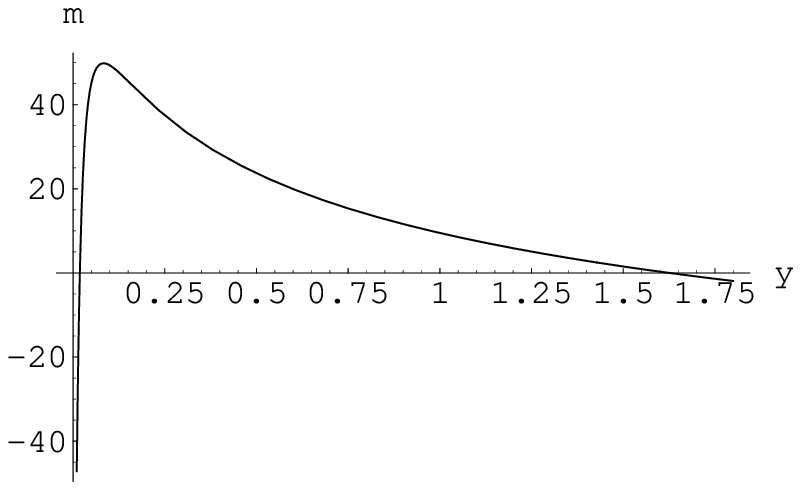}
\caption{$N$ $=$ $5$, $\f{y}{\pi} \rightarrow
\f{m_{1}(d)-m_{\textrm{\small{exact}}}}{m_{\textrm{\small{exact}}}}100$}
\label{n51loop}
\end{center}
\begin{center}
\includegraphics{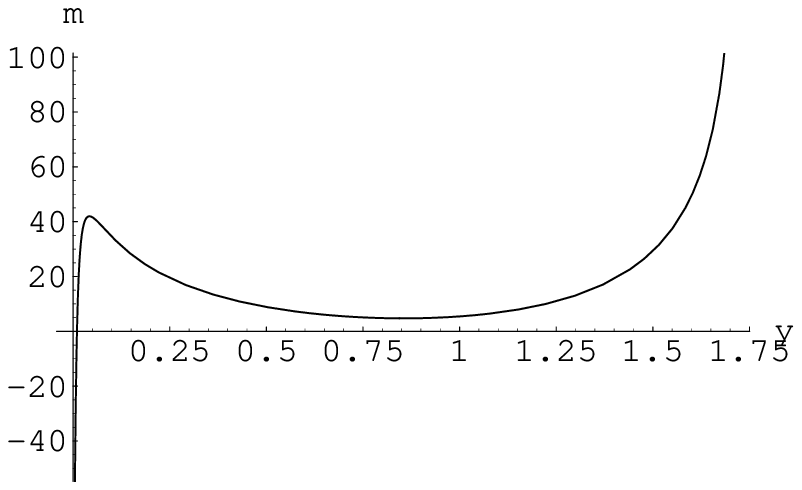}
\caption{$N$ $=$ $5$, $\f{y}{\pi} \rightarrow
\f{m_{2}(d)-m_{\textrm{\small{exact}}}}{m_{\textrm{\small{exact}}}}100$}
\label{n52loop}
\end{center}
\end{figure}

The equation for $m(d)$ (\ref{j=0}) can only be solved
numerically. If we consider instead the expansion parameter $y$ as
the free parameter one can solve it analytically to find $m(y)$.
Indeed, we can rewrite (\ref{j=0}) as \bea
&&1+y\Big(A_{0}-\f{\g{0}}{2\b{0}}k(m,y)\Big)
+y^{2}\left(B_{0}-\f{\g{0}}{2\b{0}}\left(\f{\b{2}}{\b{0}}
-\left(\f{\b{1}}{\b{0}} \right)^{2}\right) \right. \nonumber\\
&&\left. +~k(m,y)\left(A_{0}\left(1-\f{\g{0}}{2\b{0}}\right)
- \f{\g{0}\b{1}}{2\b{0}^2}\right) \right. \nonumber \\
&& \left. +~
k(m,y)^2\left(\f{\g{0}}{4\b{0}}\left(\f{\g{0}}{2\b{0}}-1\right)
\right) \right)+\ldots ~=~ 0 \label{j=02} \eea with \be
k(m,y)\equiv
d(m,y)+\f{\b{1}}{\b{0}}\ln(\b{0}y)=\f{1}{y}-\b{0}\ln\f{m^2}{\Lambda_{\l{MS}}^2}
+\f{\b{1}}{\b{0}}\ln(\b{0}y) ~. \label{km} \ee The one- and
two-loop truncation of (\ref{j=02}) is now solved easily. At one
loop it is a linear equation in $k$ and one finds
$k=\f{2\b{0}}{\g{0}}(\f{1}{y}+A_{0})$. After substituting this
into (\ref{km}) we find the one loop mass gap to be \be m_{1}(y)
~=~ \Lambda_{\l{MS}}(\b{0}y)^{\f{\b{1}}{2 \b{0}^2}} \exp \left[
\f{1}{y}\left(\f{1}{2\b{0}}-\f{1}{\g{0}}\right)-\f{A_{0}}{\g{0}}\right]
\ee The 2-loop truncation gives a quadratic equation in $k$, with
two roots $k_{1}(y),k_{2}(y)$. Hence, the two solutions for the
mass gap are \be m_{2i}(y)=\Lambda_{\l{MS}}(\b{0}y)^{\f{\b{1}}{2
\b{0}^2}}\exp
\left[\f{1}{2\b{0}}\left(\f{1}{y}-k_{i}(y)\right)\right] ~. \ee
The behavior of $m_{1}(y)$ is more or less the same for all values
of $N$. One observes a sharp maximum, followed by an asymptotic
descent to zero. There is no region of minimal sensitivity. For
$N>2$ the situation changes at two loops. One of the two solutions
$m_{2i}(y)$ has, in addition to the sharp maximum, a rather flat
minimum. This is the point of minimal sensitivity. In figures
\ref{n51loop} and \ref{n52loop} we plot the one- and two-loop
solutions for the generic $N=5$ case.

The other 2-loop solution is not physical since it varies
enormously in the region of interest, defined as the region with
acceptable estimated error, and no minimal sensitivity is found.
For $N=2$ the two-loop solution has no minimum with instead only a
rather sharp maximum at $68\%$ deviation. No true minimal
sensitivity point can be identified. The results for $N>2$ are
displayed in table \ref{minimal}. They are slightly better than
the two-loop physical scheme. Again we find II to provide a better
indication on the error then $y/\pi$. We finally remark that also
the minimal sensitivity condition can be solved exactly, to give
an analytic form of the 2-loop mass gap. We will not present it
here, however, since it is a large expression and does not give
any new insights.
\begin{table}\label{minimal}
\begin{center}
\begin{tabular}{|r|r|r|r|}
\hline $N$&$m_{2}$&$y/\pi$&II \\
\hline 3&-10.3\%&2.0&0.63\\
4&2.4\%&1.2&0.25\\
5&4.7\%&0.9&0.15\\
6&5.2\%&0.7&0.10\\
7&5.1\%&0.6&0.07\\
8&4.9\%&0.5&0.05\\
9&4.6\%&0.4&0.04\\
10&4.3\%&0.4&0.04\\
15&3.2\%&0.3&0.02\\
20&2.5\%&0.2&0.01\\
$\infty$&0\%&0&0\\\hline
\end{tabular}
\caption{Minimal sensitivity results.}
\end{center}
\end{table}

\section{Conclusions}
We have successfully applied the source inversion method to the
chiral Gross-Neveu model. This required a two-loop calculation of
the mass gap in the massive NATM which we carried out exactly.
Comparison with the exact result for the non-perturbative mass gap
gives a satisfying match. For the physical scheme, there is
convergence of the 2-loop result versus the 1-loop result. The
2-loop results are good for $N>2$ with a $10\%$-deviation for N=3
and $\leq5\%$ for $N>3$. The minimal sensitivity condition gives a
slight improvement. As in the case of the ordinary Gross-Neveu
model, the $N$~$=$~$2$ result is poor. The two-loop physical
scheme gives a $46\%$ deviation. The success/failure of the method
for $N>2/N=2$ is fairly consistent with the error estimation one
obtains from the second order term in the mass gap equation.
Finally, it would be worthwhile to apply the technique discussed
here to other models where exact mass gaps are also available.
This would have the long term aim of applying the procedure to
theories where the only information on the dynamical generated
mass comes from say Schwinger Dyson or lattice methods in order to
ascertain how competitive the results would be.

\vspace{1cm} \noindent {\bf Acknowledgement.} The two loop
calculations were performed with the use of {\sc Form},
\cite{form}.

\appendix

\sect{Computation of two loop integrals.} In this appendix we
discuss the evaluation of the basic Feynman integrals which
underly the {\em exact} value of our mass gap at two loops. At one
loop there is only one basic integral which is defined by
\begin{equation}
I ~=~ i \int_k \frac{1}{[k^2-m^2]}
\end{equation}
in Minkowski space where $\int_k$ $=$ $\int d^\omega k/(2\pi)^2$
and it has the exact value in $\omega$-dimensions
\begin{equation}
I ~=~ \frac{\Gamma(1-\omega/2)m^{(\omega-2)/2}}{(4\pi)^{\omega/2}}
~.
\end{equation}
Therefore, if $\omega$ $=$ $2$ $-$ $\epsilon$ then $I$ has a
simple pole in $\epsilon$ which is the foundation of the one loop
renormalization. At two loops all contributions to the $2$-point
function can be reduced to several basic Feynman integrals. These
are $I^2$, $\Delta(p^2)$ and $\Delta_{\mu\nu}(p^2)$ where
\begin{equation}
\Delta(p^2) ~=~ i^2 \int_k \int_l
\frac{1}{[(k-p)^2-m^2][l^2-m^2][(k-l)^2-m^2]} \label{del0}
\end{equation}
and
\begin{equation}
\Delta_{\mu\nu}(p^2) ~=~ i^2 \int_k \int_l \frac{k_{\mu}k_{\nu}}
{[(k-p)^2-m^2][l^2-m^2][(k-l)^2-m^2]} \label{del2}
\end{equation}
and these latter functions only occur in the sunset topology.
Other integrals with an obvious definition such as $\Delta_\mu
(p)$ and $\Delta_{\mu\nu\sigma}(p)$ also arise but the relevant
$2$-point function contributions can be related to (\ref{del0})
and (\ref{del2}), \cite{g90}. For instance,
\begin{equation}
\Delta_\mu(p) ~=~ \frac{2}{3} p_\mu \Delta(p^2) ~~~,~~~ p^\nu
\Delta_{\mu ~ \nu}^{~\,\mu}(p) ~=~ 2 p^\mu p^\nu
\Delta_{\mu\nu}(p^2) ~-~ \frac{2}{3} p^2 (p^2-m^2) \Delta(p^2) ~.
\end{equation}
It is elementary to observe that $\Delta(p^2)$ is finite in two
dimensions. Hence, for the mass gap we only need to evaluate it in
two dimensions when $p^2$~$=$~$m^2$. To do this we follow the
Feynman parameter approach of \cite{r89} which gives
\begin{equation}
\Delta(p^2) \,=\, \frac{\Gamma(3-\omega)}{(4\pi)^\omega} \!
\int_0^1 \! \! dx \! \int_0^1 \! \! \! dy \,
\frac{[xy(1-y)]^{1-\omega/2}}{[ y(1-y)( x(1-x)p^2 - (1-x)m^2) -
xm^2]^{3-\omega}} \label{deldefn}
\end{equation}
in $\omega$-dimensions after carrying out the momentum
integrations. Restricting to two dimensions the $y$-integration
can be performed from an integral representation of the
hypergeometric function, ${}_2F_1(a,b;c;z)$, giving
\begin{eqnarray}
\left. \Delta(p^2) \right|_{\omega=2} &=& \frac{1}{(2\pi)^2}
\int_0^1
\frac{dx}{[x(1-x)p^2 - (1+3x)m^2]} \nonumber \\
&& \times ~ {}_2F_1 \left( 1, \frac{1}{2}; \frac{3}{2};
\frac{[x(1-x)p^2-(1-x)m^2]}{[x(1-x)p^2-(1+3x)m^2]} \right) ~.
\end{eqnarray}
Next we set $p^2$ $=$ $m^2$ in the two dimensional integral to
obtain
\begin{equation}
\left. \Delta(p^2) \right|_{\omega=2 ~,~ p^2 = m^2} ~=~ -~
\frac{1}{(2\pi m)^2} \int_0^1 \frac{dx}{(1+x)^2} ~ {}_2F_1 \left(
1, \frac{1}{2} ; \frac{3}{2}; \frac{(1-x)^2}{(1+x)^2} \right)
\end{equation}
which reduces to
\begin{equation}
\left. \Delta(p^2) \right|_{\omega=2 ~,~ p^2 = m^2} ~=~
\frac{1}{8\pi^2m^2} \int_0^1 dx \, \frac{\ln x}{(1-x^2)} ~.
\end{equation}
The final integral can now be calculated exactly, \cite{gr65}, to
produce
\begin{equation}
\left. \Delta(p^2) \right|_{\omega=2 ~,~ p^2 = m^2} ~=~ -~
\frac{3\zeta(2)}{32\pi^2m^2} ~. \label{delres}
\end{equation}

The remaining integral which arises in the sunset topology occurs
with two Lorentz contractions. First, in $\omega$-dimensions
without setting the on-shell condition it is straightforward to
show that, \cite{g90},
\begin{equation}
\Delta_\mu^{~\mu}(p^2) ~=~ I^2 ~+~ \frac{1}{3} (p^2 + 3m^2)
\Delta(p^2) ~.
\end{equation}
Although the contraction of (\ref{del2}) with $p_\mu p_\nu$ is
also divergent it cannot be written in a similar closed form.
However, its divergent part is known to be $p^2 I^2/\omega$,
\cite{g90}. Therefore,
\begin{equation}
F_\Delta(p^2) ~=~ \omega p^\mu p^\nu \Delta_{\mu\nu}(p^2) ~-~
p^2\Delta_\mu^{~\mu}(p^2)
\end{equation}
will be finite in two dimensions and can be evaluated exactly when
the on-shell condition is set similar to the derivation of
(\ref{delres}). With the same Feynman parametrization as
(\ref{deldefn}) we have
\begin{eqnarray}
F_\Delta(p^2) &=&
\frac{(\omega-1)\Gamma(3-\omega)(p^2)^2}{(4\pi)^\omega}
\nonumber \\
&& \times \int_0^1 dx \int_0^1 dy
\frac{(1-x)^2[xy(1-y)]^{1-\omega/2}}{[
y(1-y)( x(1-x)p^2 - (1-x)m^2) - xm^2]^{3-\omega}} ~. \nonumber \\
\end{eqnarray}
Hence, using the properties of the hypergeometric function again
we find
\begin{equation}
\left. F_\Delta(p^2) \right|_{\omega=2 ~,~ p^2 = m^2} ~=~
\frac{m^2}{8\pi^2} \int_0^1 dx \, \frac{(1-x)}{(1+x)} \ln x
\end{equation}
leading to, \cite{gr65},
\begin{equation}
\left. F_\Delta(p^2) \right|_{\omega=2 ~,~ p^2 = m^2} ~=~
\frac{m^2}{8\pi^2} [ 1 ~-~ \zeta(2) ] ~. \label{del2res}
\end{equation}
Hence, all integrals in the full $2$-point function can be written
in terms of $I^2$, (\ref{delres}) and (\ref{del2res}).

We have checked that the values obtained here for the finite parts
of $\Delta(p^2)$ and $\Delta_{\mu\nu}(p^2)$ agree with the
numerical values given in \cite{agikn97}. For instance, repeating
the calculation which leads to equation (4.5) of \cite{agikn97} we
find that $[0.737775$~$-$~$\pi^2/96]$ corresponds to
$[\zeta(2)/2$~$-$~$3/16]$.

\bibliographystyle{unsrt}

\end{document}